\documentclass[aps,prb,twocolumn,superscriptaddress,reprint]{revtex4-1}

\usepackage{bm}
\usepackage{graphicx}
\usepackage{latexsym}
\usepackage{amssymb}
\usepackage{amsmath}

\begin{document}

\title{Manipulation of electronic and magnetic properties of M$_2$C (M=Hf, Nb, Sc, Ta, Ti, V, Zr) monolayer by applying mechanical strains}
\author{\surname{Shijun} Zhao}
\author{\surname{Wei} Kang}
\email{weikang@pku.edu.cn}
\affiliation{HEDPS, Center for Applied Physics and Technology, Peking University, Beijing 100871, P. R. China}
\affiliation{College of Engineering, Peking University, Beijing 100871, P. R. China}
\author{\surname{Jianming} Xue}
\affiliation{HEDPS, Center for Applied Physics and Technology, Peking University, Beijing 100871, P. R. China}
\affiliation{State Key Laboratory of Nuclear Physics and Technology, School of Physics,
             Peking University, Beijing 100871, P. R. China}
%%%%%%%%%%%%%%%%%%%%%%%%%%%%%%%%%%%%%%%%%%%%%%%%%%%%%%%%%%%

\begin{abstract}

Tuning the electronic and magnetic properties of a material through strain engineering is an effective strategy to enhance the performance of electronic and spintronic devices. Recently synthesized two-dimensional transition metal carbides M$_2$C (M=Hf, Nb, Sc, Ta, Ti, V, Zr), known as MXenes, has aroused increasingly attentions in nanoelectronic technology due to their unusual properties. In this paper, first-principles calculations based on density functional theory are carried out to investigate the electronic and magnetic properties of M$_2$C subjected to biaxial symmetric mechanical strains. At the strain-free state, all these MXenes exhibit no spontaneous magnetism except for Ti$_2$C and Zr$_2$C which show a magnetic moment of 1.92 and 1.25 $\mu_B$/unit, respectively. As the tensile strain increases, the magnetic moments of MXenes are greatly enhanced and a transition from nonmagnetism to ferromagnetism is observed for those nonmagnetic MXenes at zero strains. The most distinct transition is found in Hf$_2$C, in which the magnetic moment is elevated to 1.5 $\mu_B$/unit at a strain of 1.80\%. We further show that the magnetic properties of Hf$_2$C are attributed to the band shift mainly composed of Hf(5$d$) states. This strain-tunable magnetism can be utilized to design future spintronics based on MXenes.
\end{abstract}
%%%%%%%%%%%%%%%%%%%%%%%%%%%%%%%%%%%%%%%%%%%%%%%%%%%%%%%%%%%

\maketitle

%%%%%%%%%%%%%%%%%%%%%%%%%%%%%%%%%%%%%%%%%%%%%%%%%%%%%%%%%%%
Two-dimensional (2D) materials have attracted a great deal of attention due to their unique mechanical and electronic properties and numerous potential applications.\cite{Novoselov2004} Very recently, a family of 2D materials was synthesized by the exfoliation of the layered ternary transition metal carbides, which are known as MAX phase.\cite{Barsoum2000,Eklund2010} The MAX phases represent a large (more than 60 members) family of ternary layered machinable transition metal carbides, nitrides, and carbonitrides, which possess unique properties such as remarkable machinability, high damage tolerance, excellent oxidation resistance, and high electrical as well as thermal conductivity.\cite{Barsoum2011,Wang2010} The MAX phases are layered hexagonal (space group P63/$mmc$) with the formula of M$_{n+1}$AX$_n$ (n=1,2,3), where ¡§M¡¨ represents an early transition metal, ¡§A¡¨ is an A-group element (mostly groups IIIA and IVA) and ¡§X¡¨ denotes carbon or nitrogen. The layered structure makes the exfoliation of MAX phase feasible and by removal of the ¡§A¡¨ group layer from the MAX phase, 2D MXenes can be obtained which share the structural similarities with graphene.\cite{Naguib2011,Naguib2012}

The extraordinary properties of 2D MXenes have been intensely investigated since its successfully synthesization experimentally.\cite{Enyashin2012,Naguib2012a,Tang2012,Mashtalir2013,Khazaei2013,Lane2013,Naguib2012,Naguib2011,Come2012} High conductivities and high elastic moduli are found in MXenes.\cite{Naguib2011,Naguib2012,Enyashin2012} Besides, the applications of MXenes in Li-ion battery anodes and hybrid electro-chemical capacitors are also demonstrated.\cite{Tang2012,Naguib2012a} Most of the synthesized M$_{n+1}$X$_n$ layers are metallic with no band gap, although it is shown that the electronic structure can be modulated by appropriate surface terminations such as oxygen and fluoride.\cite{Naguib2011,Enyashin2013} However, previous theoretical studies indicate that the majority of the functionalized MXene systems are nonmagnetic. For the ``pure'' MXenes, only the Ti$_{n+1}$X$_n$ (X=C,N) monolayers are magnetic due to the  3$d$ electrons of surface Ti atoms.\cite{Xie2013}

In the applications based on nanoelectronic devices, an essential requirement is the versatile ability to modulate the electronic and magnetic properties via external controls. It has been well established that the electronic and magnetic properties of materials can be tuned by applying mechanical strains.\cite{Yun2012,Peelaers2012} In particular, the effects of strain will be more significant in 2D systems as demonstrated in graphene and other transition metal chalcogenides such as MoS$_2$, NbS$_2$, NbSe$_2$ and VSe$_2$.\cite{Zhou2012,Yun2012} For example, a uniaxial strain larger than 20\% applied in graphene along the zigzag direction is predicted to open a bandgap due to level crossing.\cite{Pereira2009}Different from bulk materials in which strains can be relaxed by dislocation plasticity or fracture, 2D materials are well suited for strain engineering. From experimental view point, strain in 2D sheet can be achieved rather easily by manipulating its substrate.

In the present work, therefore, we study the effect of mechanical strains on the electronic properties of MXenes of the form M$_2$C (M=Hf, Nb, Sc, Ta, Ti, V, Zr). The influence of homogeneous biaxial tensile strain on these MXenes are systematically investigated based on density functional theory calculations. Our results demonstrate that the Hf$_2$C structure can be efficiently magnetized by simply applying a tensile strain, during which the structures of MXenes are still maintained. The effect of compressive strain is not considered in the present work because it is impractical for experimental implementation.

Electronic structure calculations have been carried out using the density functional theory as implemented in the Vienna $ab~initio$ simulation (VASP) code.\cite{Kresse1996} The electron-ion interactions are described by Projector-augmented-wave (PAW)  pseudopotentials with semi-core states treated as valence states.\cite{Blochl1994} The exchange and correlation energies are treated within the generalized gradient approximation (GGA) according to PW91 parametrization.\cite{Perdew1992} To obtain the unstrained configuration of M$_2$X monolayer, both the atomic positions and lattice vectors are fully relaxed with the conjugate gradient algorithm to obtain the equilibrium lattice constant. The lattice constant is then increased gradually to model the biaxial strain, in which all atoms are optimized to calculate the electronic structure. The convergence for energy is chosen as 10$^{-4}$ eV between two consecutive self-consistent steps and the atomic relaxation is performed until the interatomic forces are less than 0.01 eV/\AA. A plane wave cutoff of 520 eV is used for all the calculations.

In our calculations, the M$_2$X is represented by a 4$\times$4 supercell with a vacuum space of 18~\AA~between two adjacent layers to avoid interactions between periodic images of slabs in the $z$-direction. The Brillouin zone is sampled using a 6$\times$6$\times$1 Monkhorst-Pack grid in relaxation calculations, while a denser mesh of 12$\times$12$\times$1 is considered to generate density of states and charge densities.

\begin{figure}[!htb]
   \begin{center}
   \includegraphics[width=0.4\textwidth]{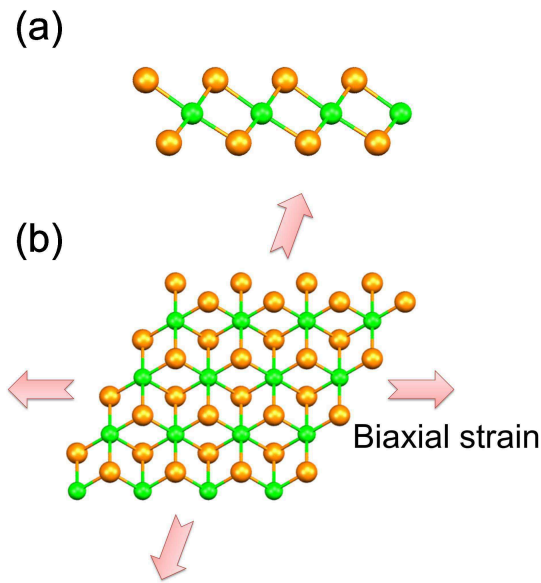}
  \end{center}
   \caption{(a) Side and (b) top view of M$_2$C, where the M and X atoms are denoted by orange and yellow spheres, respectively.}
   \label{configure}
\end{figure}

The side and top views of generalized structure of M$_2$X MAxenes with hexagonal symmetry are illustrated in Fig. \ref{configure}, which are constructed by removing ``A'' atoms from corresponding bulk MAX phases. In Fig. \ref{configure}(b), the schematic illustration of the biaxial tensile strain are indicated. Here, the strain is defined as $\epsilon=(a-a_0)/a_0=\Delta a/a_0$, where $a_0$ and $a$ is the lattice parameter of the unstrained and strained supercell, respectively. As described previously, the structure of M$_2$X suggest that it can bear a relatively large range of in-plane elastic strains compared to the perfectly flat graphene and BN layer due to the possible deformation in the perpendicular direction.\cite{Zhou2012}

%++++++++++++++++++++++++++++++++++++++++++++++++++++++++++++++++++++++
%++++++++++++++++++++++++++++++++++++++++++++++++++++++++++++++++++++++

To study the effect of strain on the electronic properties of MXenes, we first relaxed the atomic positions as well as lattice vectors and obtained their optimized structures. The calculated lattice parameters of M$_2$X monolayer is summarized in Table \ref{Tab1} with available data from literatures shown for a comparison.

\begin{table}[!htb]
    \caption {Lattice parameters (in \AA) of M$_2$X monolayer determined with PW91 functional in this work. Available results from literatures computed from different functionals are also included for a comparison.}
	\centering
		\begin{tabular} {c  c  c  c  c  c  c  c}
			\hline\hline
		        MXenes      & Hf$_2$C & Nb$_2$C  & Sc$_2$C & Ta$_2$C & Ti$_2$C &  V$_2$C &  Zr$_2$C \\
				\hline
			    present     & 3.212   &  3.135   & 3.332   & 3.087   &  3.078  &  2.897  &  3.293 \\
        PBE\cite{Gan2013}   &         &          &         &         &  3.076  &         &        \\
        PBE\cite{Shein2012} &         &          &         &         &  3.0395  &        &        \\
    DFTB\cite{Enyashin2013a}&         &          &         &         &  3.33  &        &        \\
 Wu-Cohen\cite{Kurtoglu2012}& 3.239   &          &         & 3.138   &  3.007   & 2.869   & 3.238 \\
            \hline\hline
  		\end{tabular}
  \label{Tab1}
\end{table}

It can be seen that our results are in good agreement with available theoretical calculations, which also validates our approach in predicting the structural properties of these MXenes. Note that there is still no experimental values about the lattice parameters. Here it is helpful to make a comparison of these lattice constants with their corresponding MAX phases. Taking Ti$_2$AlC as an example, the lattice constant $a$ of bulk Ti$_2$AlC is 3.065~\AA~ determined from experiment,\cite{Wang2010}, which is comparable to the calculated constant of Ti$_2$C as shown above. Therefore, it can be concluded that the structures of MXenes are almost the same as those in MAX phases.

\begin{figure}[!htb]
   \begin{center}
   \includegraphics[width=0.45\textwidth]{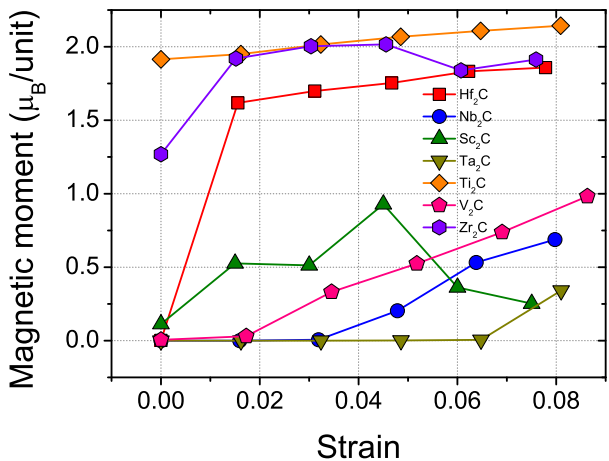}
  \end{center}
   \caption{Dependence of magnetic moments of various MXenes on the applied strain.}
   \label{moment}
\end{figure}

After the stable structures of MXenes are obtained, we start to apply biaxial symmetric strains to the lattice to investigate the influence of strain on their electronic and magnetic properties. Previous theoretical calculations based on density functional theory have revealed that most of the MXenes are metallic. Indeed, our results demonstrate that all the investigated MXenes in this study are metallic system with no energy gap. This is still the case when strain is applied and the metallic properties of these MXenes are not changed. However, the magnetic properties are much affected. Here the total magnetic moment of MXenes as a function of applied strains are presented in Fig. \ref{moment}.

As we can see, the magnetic properties of these MXenes can be effectively tuned by external strains. At the strain-free state, only the Ti$_2$C and Zr$_2$C has magnetic moment, which is 1.92 and 1.25 $\mu_{B}$/unit respectively. The results for Ti$_2$C is in good agreement with previous calculations of 1.85 $\mu_B$/unit.\cite{Xie2013} A noteworthy feature is that the Hf$_2$C display no magnetism although it is in the same group as Ti and Zr in the periodic table. In fact, it has been reported that in the MAX phase M$_2$AC, this group elements (M=Ti, Zr and Hf) exhibit the lowest bulk moduli than other MAX phases. The reason is interpreted as the weak M $d$-orbital and C $p$-orbital bonding in these materials.\cite{Warner2006} As the atomic number increases, more carbon $p$ electrons are participated in the M-C bonding. In the case of MXenes, the same trend is observed. The $p$ orbital of carbon atom hybridizes weakly with $d$ orbital of Ti, while a strong hybridization is observed between Hf(5$d$) and C(2$p$) orbitals. Therefore, the unsaturated Ti(3$d$) dangling bonds lead to the magnetism in the Ti$_2$C monolayer. On the contrary, the strong bonding interaction between Hf and C gives rise to no magnetism in Hf$_2$C, which will be discussed in more details below.

All the magnetic moment of these MXenes are elevated by applied strains, although they show different behaviors. The most distinct variation is observed for Hf$_2$C, which turns from nomagnetism to ferromagnetism with a high magnetic moment of 1.70 $\mu_B$/unit at a strain of 2\%. In fact, the system becomes magnetic as long as the strain is applied, with a linearly increase of the magnetic moment. At about 6\% strain, the spin moment increases slowly and remains almost constant for larger strains. This indicates that a slight variation of the strain applied on the Hf$_2$C layer can induce a large change of the spin moment.

%++++++++++++++++++++++++++++++++++++++++++++++++++++++++++++++++++++++
%++++++++++++++++++++++++++++++++++++++++++++++++++++++++++++++++++++++

\begin{figure}[!htb]
   \begin{center}
   \includegraphics[width=0.38\textwidth]{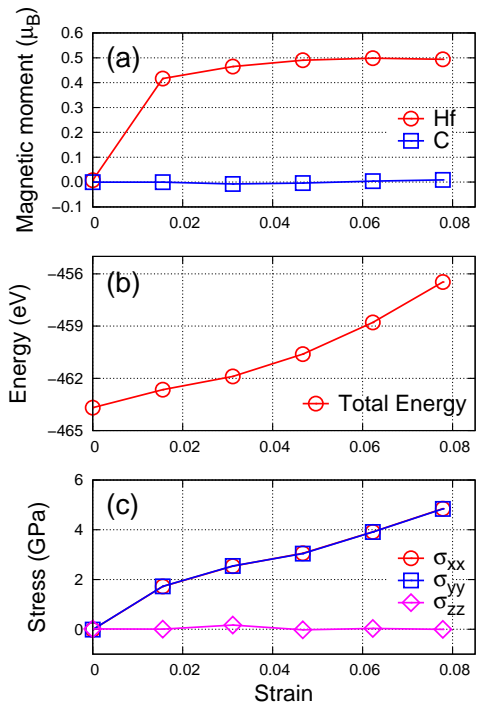}
  \end{center}
   \caption{Per atom magnetic moment, total energy and stress of Hf$_2$C as a function of applied strain.}
   \label{pmag}
\end{figure}

As a prototype of strain-engineered modulation on the magnetic properties of MXenes, we have made a detailed analysis on the interplay between the mechanical strain and the magnetic behaviors for Hf$_2$C. The magnetic moments of Hf and C atoms as a function of the applied strain is represented in Fig.~\ref{pmag}(a). It is shown that the magnetic moment of Hf starts to increase sharply when the strain is applied and saturates to a value of 0.5 $\mu_B$/atom at a strain of 5\%. The increase of magnetic moment in the strain range up to 2\% is nearly linear with respect to the strain, which means that the magnetic property of Hf$_2$C is rather sensitive to the strain. On the other hand, the nonmagnetic property of C is not much influenced by the strain. These results suggest that the magnetism of Hf$_2$C under strains is mainly originated from Hf atoms.

%++++++++++++++++++++++++++++++++++++++++++++++++++++++++++++++++++++++
%++++++++++++++++++++++++++++++++++++++++++++++++++++++++++++++++++++++

In Fig.~\ref{pmag} (b) and (c), we have shown the dependence of the total energy and stress of Hf$_2$C sheet on the applied strain. The total energy is monotonically increased with strain, indicating the structure is still stable when it is subjected to these strains since no phase transition is observed. This implies that the Hf$_2$C monolayer can bear a large elastic strain range. The smoothly increase also suggests that the magnetic modulation is in the elastic range and reversible. This feature is significant for practical nanomechanical control of the properties of the MXenes system in device applications. The in-plane stresses of $\sigma_{xx}$ and $\sigma_{yy}$ are almost the same, which is the result of biaxial symmetric strains.

\begin{figure}[!htb]
   \begin{center}
   \includegraphics[width=0.48\textwidth]{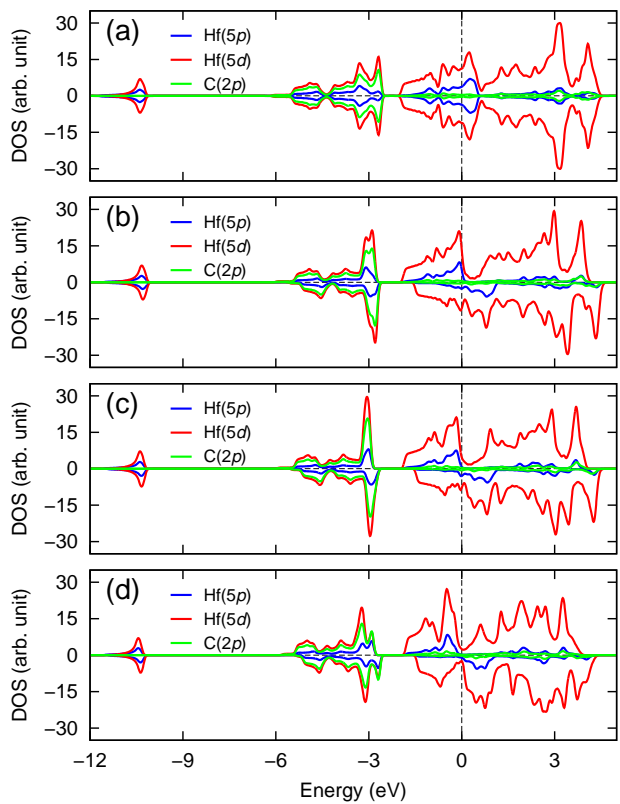}
  \end{center}
   \caption{Evolution of PDOS of Hf$_2$ along with biaxial symmetric strain: (a) 0.00\%, (b)1.56\%, (c)4.67\% and (d)7.78\%.
            }
   \label{pdos}
\end{figure}

It is well established that the partially filled $d$ character of the transition metal atoms can induce significant magnetism in materials. In our case, the magnetism is indeed induced by the Hf atom as demonstrated in Fig.~\ref{pmag}. The Hf atom possesses two 5$d$ electrons. However, due to the strong covalent interactions between Hf and C in the sandwiched structures, the monolayer MXene of Hf$_2$C exhibits no magnetism. This phenomena have been observed in other monolayer structures such as NbS$_2$ and NbSe$_2$. In these materials, the strong ligand field owing to the covalent bonds tend to quench the magnetism of the transition metal. When external strain is applied, the covalent bonding interactions inside Hf$_2$C is reduced and result in the release of the feature of $d$ electrons of Hf and finally give rise to magnetism in the system. In order to analyze the origin of the magnetism observed above, we show in Fig.~\ref{pdos} the projected density of states (PDOS) of Hf$_2$C monolayer under different strains.
%++++++++++++++++++++++++++++++++++++++++++++++++++++++++++++++++++++++
%++++++++++++++++++++++++++++++++++++++++++++++++++++++++++++++++++++++

It can be seen from Fig.~\ref{pdos}(a) that the pristine Hf$_2$C is a metallic system where the energy states near the Fermi energy are mainly composed of Hf(5$d$) orbitals. In the energy range from -5 to -2 eV, the states of C(3$p$) and Hf(5$d$) hybridize, forming bands separated by a gap of around 1eV from the Hf($d$) bands. The metallic nature of Hf$_2$C is not altered by applied strain since there are always bands crossing the Fermi level. However, the split of band is induced. The spin-up states of Hf(5$d$) and Hf(5$p$) start to shift towards lower energies, while the spin-down states shift towards higher energies. As a result, the magnetism is observed which is mainly contributed by the Hf(5$d$) states. This magnetism is ascribed to the redistribution of atomic charges under external strain.

Closer examination of the PDOS reveals that the $d_{yz}$ and $d_{xz}$ states as well as $d_{xy}$ and $d_{x^2-y^2}$ states of Hf are degenerated. The largest contribution for the magnetism is from the shift of $d_{yz}$ and $d_{xz}$ bands. This is in accordance with the biaxial strain we applied. Since the strain is symmetric, the degeneracy is not destroyed.

\begin{figure}[!htb]
  \begin{center}
  \includegraphics[width=0.35\textwidth]{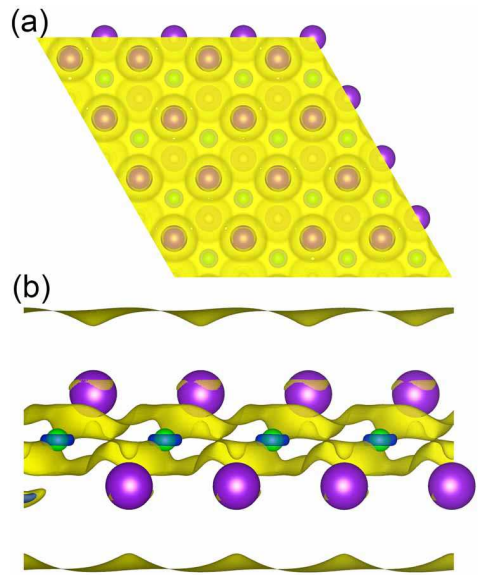}
  \end{center}
  \caption{The spin density distribution ($\rho\uparrow$ - $\rho\downarrow$) with the positive portion denoted by yellow and negative denoted    by blue. The isovalue is set to be 1 $\times$ 10$^{-4}$ e/\AA$^3$.
  }
   \label{spndes}
\end{figure}

The spin density plot at a strain of 1.56\% is provided in Fig.~\ref{spndes}. At strain free state, there is no spin density in the Hf$_2$C monolayer, which is the refection of its nonmagnetic property. It is shown that the strain induced ferromagnetism is mainly attributed the Hf since the unpaired spin electrons are mostly concentrated on the Hf atoms. Only a small fraction of spin density is located at C atoms. This is consistent with the results shown in the PDOS in Fig.~\ref{pdos}.

%++++++++++++++++++++++++++++++++++++++++++++++++++++++++++++++++++++++
%++++++++++++++++++++++++++++++++++++++++++++++++++++++++++++++++++++++

\begin{figure}[!htb]
   \begin{center}
   \includegraphics[width=0.45\textwidth]{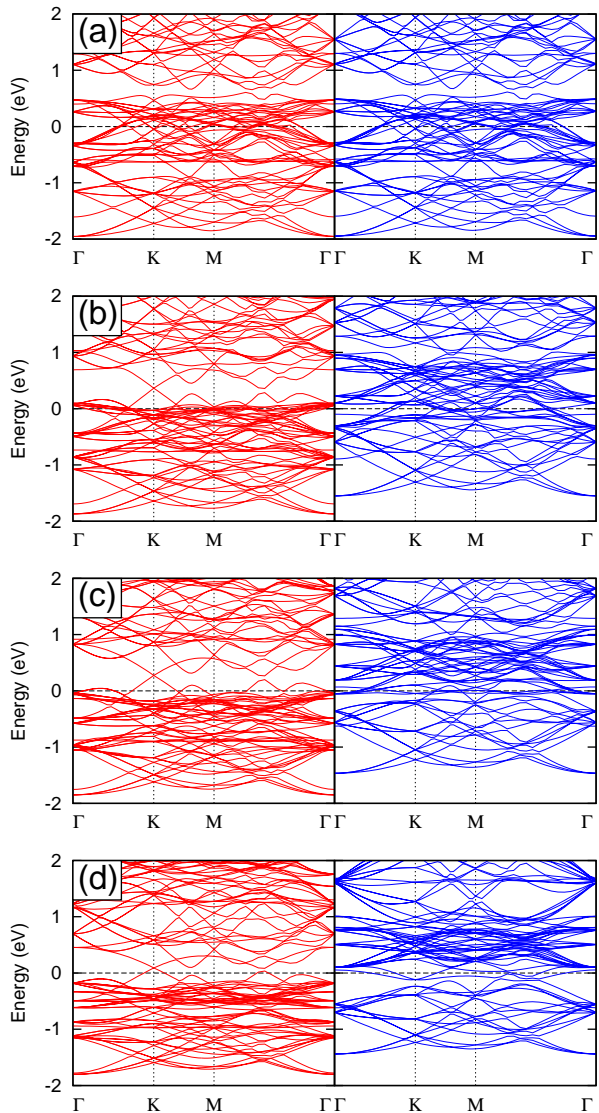}
  \end{center}
   \caption{Spin-resolved band structure of Hf$_2$ under different biaxial strains: (a) 0.00\%, (b)1.56\%, (c)4.67\% and (d)7.78\%.
            }
   \label{band}
\end{figure}

The spin-resolved band structure of Hf$_2$C under different biaxial strains is shown in Fig.~\ref{band}, which clearly displays the opposite shift of Fermi level in the majority and minority spin bands when the strain is applied. As a consequence, the symmetric band structure becomes asymmetrical due to the shift of conduction and valence bands, in accordance with the DOS plot. The energy bands in the vicinity of Fermi level determines the magnetic behaviors as demonstrated above. The band splitting in this region changes Hf$_2$C to a magnetic metal.

%++++++++++++++++++++++++++++++++++++++++++++++++++++++++++++++++++++++
%++++++++++++++++++++++++++++++++++++++++++++++++++++++++++++++++++++++
The effects of spin-orbit coupling (SOC) on the calculated magnetism are also evaluated. With the unit cell of Hf$_2$C, we performed SOC calculations which treat magnetism noncollinearly using quantization axis (0, 0, 1) (i.e., $z$ axis) based on the charge density from no SOC. Although the inclusion of SOC lowers the total energy about 1.42 eV, the calculated magnetic moment is not influenced. This result is in line with previous calculations regarding the effects of SOC on magnetic moment.\cite{Xie2013a} Therefore, our conclusion still holds irrespective of SOC.

We also checked the dependence of our results on the functional we have used so as to give more reliability of our results. In this case, we have performed a set of calculations about Hf$_2$C with PBE functional which has frequently used for the study of MXenes.\cite{Gan2013,Shein2012} The results indicate that the dependence of magnetic moments of Hf$_2$C on applied strain is more sensitive for PW91 than PBE functional at small strains. Nevertheless, both functionals predict a non-magnetic to magnetic transition when the strain is larger than 2\%. Therefore, the main results discussed in this paper are not influenced.

Since the experimental synthesized MXenes are usually terminated with oxygen-containing and/or fluoride groups, we also examine the effect of strain on their electronic structures. The results show that all the MXenes considered exhibit no magnetism upon fictionalization. This conclusion is also found in Ti$_2$C in previous calculations.\cite{Xie2013} Our further simulation indicates that the nonmagnetic properties of H-, O- and F- functionalized Ti$_2$C monolayer is not changed along with applied strains.

In 2D materials, a conventional method to induce magnetism is to introduce vacancy or foreign atoms.\cite{Lehtinen2003,Santos2008} However, the precise control of these structure is rather difficult experimentally. Besides, the generated magnetism in this way is often localized. Our results indicates that the magnetism of MXenes can be switched on and off facilely by applied strains. In experiment, these strains can be achieved in a controlled manner, as demonstrated recently in 2D materials such as graphene.\cite{PhysRevB.79.205433} At strain-free state, only the Cr$_2$C are found to be magnetic among various MXenes, our results thus broaden the applicability of MXenes in future spintronics.

In summary, the electronic and magnetic properties of recently synthesized MXenes subjected to external biaxial symmetric strain are investigated by virtue of first-principles calculations based on density functional theory. All the considered MXenes are metallic in the strain-free state and the metallic characteristics are not influenced even with a large applied strains. Nevertheless, the magnetic moments of these MXenes are very sensitive to strains. Among the studied MXenes, the magnetic moment of Hf$_2$C exhibits a large variation which renders it magnetic with a magnetic moment of 1.5 $\mu_B$/unit at a strain of 1.80\%. Our analysis reveals that the magnetism is originated from the shift of Hf(5$d$) bands. The present work provides a route to harness the magnetic properties of two-dimensional MXenes for spintronic applications.

\begin{acknowledgments}
This work is financially supported by the NSFC (Grant No. 11274019).
\end{acknowledgments}

%merlin.mbs apsrev4-1.bst 2010-07-25 4.21a (PWD, AO, DPC) hacked
%Control: key (0)
%Control: author (72) initials jnrlst
%Control: editor formatted (1) identically to author
%Control: production of article title (-1) disabled
%Control: page (0) single
%Control: year (1) truncated
%Control: production of eprint (0) enabled
%

%%%%%%%%%%%%%%%%%%%%%%%%%%%%%%%%%%%%%%%%%%%%%%%%%%%%%%%%%%%
%%%%%%%%%%%%%%%%%%%%%%%%%%%%%%%%%%%%%%%%%%%%%%%%%%%%%%%%%%%
\end{document}